\documentclass[acmsmall,screen, nonacm]{acmart} 
\AtBeginDocument{%
  }

\usepackage{listings}
\usepackage{xcolor}

\newcommand{\mytab}{Tab.\@~}
\newcommand{\mysec}{Sec.\@~}
\newcommand{\myfig}{Fig.\@~}

\newcommand{\peak}{\textsc{Peak}}

\usepackage{tikz}
\usetikzlibrary{shapes.geometric, positioning,fit,arrows.meta,backgrounds}

\usepackage{xspace}
\usepackage{multirow}
\usepackage{comment}
\usepackage{wrapfig}

\lstdefinestyle{terminal}{
backgroundcolor=\color{white},
basicstyle=\ttfamily\color{black},
breakatwhitespace=false,
breaklines=true,
captionpos=b,
keepspaces=true,
showspaces=false,
showstringspaces=false,
showtabs=false,
tabsize=2
}
\begin{document}

\title{PEAK: A Performance Engineering AI-Assistant for GPU Kernels Powered by Natural Language Transformations}

\author{Muhammad Usman Tariq}
\affiliation{%
 \institution{Stanford University}
 \country{USA}
}

\author{Abhinav Jangda}
\affiliation{%
 \institution{Microsoft Research Redmond}
 \country{USA}
}

\author{Angelica Moreira}
\affiliation{%
 \institution{Microsoft Research Redmond}
 \country{USA}
}

\author{Madan Musuvathi}
\affiliation{%
 \institution{Microsoft Research Redmond}
 \country{USA}
}
\author{Tyler Sorensen}
\affiliation{%
 \institution{Microsoft Research Redmond and University of California Santa Cruz}
 \country{USA}
}





\begin{abstract}
Advancements in large language models (LLMs) are showing promising impact in software development and programming assistance. However, these models struggle when operating on low-level backend code. This challenge is exacerbated in the domain of GPU kernels, where performance-critical details are coupled to rapidly evolving hardware characteristics and available code examples are sparse.

In this work, we introduce \peak{}, a Performance Engineering AI-Assistant for GPU Kernels powered by natural language transformations. \peak{} utilizes the key insight that iterative code transformations (optimizations) can straightforwardly be written in natural language, and then carried out by LLMs. Thus, these transformations can be rapidly developed, encoding general portable optimizations, but also easily specialized to specific GPU devices and even kernels. These natural transformations are supported by a modular and extensible infrastructure that additionally performs validation and performance evaluation. We demonstrate the flexibility of \peak{} by instantiating it for three backends, CUDA, HIP, and HLSL, and create 16 natural transformations for optimizing matrix multiplication kernels. We show that our resulting implementations are competitive with vendor libraries when available, and for HLSL (without a library) our implementations match the hardware documented FLOPS. \peak{} allows the fine-grained exploration of several research questions around how LLMs behave in this domain, including characterizing transformations and their errors; and how performance evolves along optimization sequences. \peak{} provides an interface that can either be utilized by performance engineers to improve productivity, or driven completely autonomously (e.g., by an AI agent), providing a forward-compatible design that can continue to improve with advances in AI capabilities.

\end{abstract}




\maketitle

\section{Introduction}


Advances in large language models (LLMs) have significantly impacted software development, assisting, or even fully automating, a range of programming tasks~\cite{githubcopilot2025, amazon2025qdeveloper}. Due to the abundance of high-level code examples and mature frameworks, LLMs have been most effective in front-end and application-level domains, while low-level backend programming remains a greater challenge~\cite{AnthropicEconomicReport}. Nonetheless, given the potential of LLMs to improve productivity and democratize software development, extending AI assistance to lower-level code is a promising research direction. GPU kernels are a natural target: they are notoriously difficult to write and tune, yet have enormous performance implications, particularly in the context of modern AI workloads. Even modest reduction in kernel runtime can yield significant gains in cost efficiency and energy consumption, especially at scale.

Despite efforts to provide more accessible GPU programming languages~\cite{tillet2021triton,tilelang:2025}, development of efficient GPU kernels remains a domain reserved for experts, requiring deep understanding of complex memory and concurrency hierarchies, synchronization models, and architecture-specific optimizations. Compilers for these programming languages necessarily aim to be general-purpose, while maximizing performance requires performing optimizations that are specific to the kernel, the particular hardware backend, and possibly specific sizes of the inputs. Even if specialized compilers are engineered to optimize to the peak performance for some kernels, the GPU ecosystem evolves rapidly, with new architectural features required to fully exploit hardware capabilities, making it very difficult for these frameworks to keep up. However, performance engineering efforts tend to follow some structure; if this structure can be captured, especially utilizing new AI capabilities, then there may be opportunities to impact this domain.

Recognizing this potential, there have been several prior (and ongoing) efforts to generate optimized GPU kernels using AI, spanning academic work, large tech companies, and startups. For example, KernelBench~\cite{kernelbench} and the AI CUDA Engineer~\cite{AICUDAEngineer} explores optimizing CUDA kernels in PyTorch using LLMs. NVIDIA has demonstrated how LLMs can be used to optimize attention kernels~\cite{nvidiallmattention}, and several startups appears to be building a product around this use case, e.g., see ~\cite{Stanley2025_Mako, su2025cudal2surpassingcublasperformance, standardkernel_2025}. While these efforts introduce promising techniques for transforming GPU code, they share a critical limitation: most adopt an all-or-nothing and opaque approach to automation, with minimal support for human-AI collaboration, interpretable iterative refinement, or extensible modular interfaces. This rigidity has led to brittle behavior and unintended side effects, in two cases, the system made critical mistakes which were not caught by the testing infrastructure~\cite{miru_why_2025_tweet, wiggers2025sakana}. 

\subsection{PEAK: Optimizing GPU Kernels through Natural Language Transformations}

This paper presents \peak{}, a GPU kernel optimization framework, powered by natural language transformation specifications executed by LLMs, to assist performance engineers. \peak{} captures the essence of expert performance-engineering processing with the following components:  

\paragraph{Natural Transformations.} Performance engineers deploy a set of optimization strategies, each expressed informally in natural language (either explicitly as documentation or implicitly in their thinking). \peak{}'s core contribution is to embrace this concept through \emph{natural transformations}. These natural transformations are written in natural language and range from general strategies, such as "unroll a loop" to precise directives, such as "tile the inner loop over the K dimension," that are specific to the kernel being optimized. Because natural transformations are simple to write (e.g., as compared to formal compiler passes), they can be highly specialized to a specific kernel.

\paragraph{Kernel Context.} A kernel context consists of a GPU kernel, host-side code that launches the kernel, input sizes, and a set of performance tuning parameters. This provides sufficient scope to perform aggressive optimizations (matching the performance of handwritten libraries for complex kernels), while also checking correctness and analyzing performance. Natural transformations can transform one kernel context into a new kernel context using LLMs.

\paragraph{Correctness Validators and Performance Evaluators.}
GPU kernels are complex, as they are executed by many threads across a complex hierarchy with features that are often poorly documented; errors are common, even when programmed by experts, let alone generated by an LLM. Thus, any natural transformation should be rigorously checked. Thus, \peak{} provides an interface for \emph{correctness validators} to check the functional correctness of a kernel context. The base correctness validator simply compares kernel output values to a baseline. Moreover, \peak{} exposes an extensible interface for adding backend-specific validators, such as the compute sanitizers from NVIDIA. In the case of a failure, a previous kernel context can be restored; from this, the LLM could simply retry, or fall back to human-in-the-loop debugging, enabling developers to refine or rephrase natural transformations.

Similar to the correctness validators, the performance evaluators operate on a kernel context. These provide information about the performance, which can be attached to the kernel context and used to drive decisions about further optimizations. The basic performance evaluator simply samples tuning parameters and checks the runtime of the kernel. However, these are extensible and we illustrate this by incorporating more advanced tools, including the OpenTuner~\cite{opentuner} autotuning framework and NVIDIA's Nsight profiler.

\paragraph{Performance Workflows and Portability.} Performance engineers iteratively apply transformations using a mix of intuition, experimentation, and domain expertise. \peak{} provides a structured way to capture this exploration with its \textit{performance workflow} abstraction. Users string together a sequence of natural transformations in a way that resembles \texttt{git} for software versioning. Every time that a natural transformation modifies a kernel context, it can be analyzed for correctness and performance; upon successful completion, it creates a checkpoint in the performance workflow. These checkpoints can be named, compared, restored, and revisited. As users dynamically explore their optimization process, the performance workflow and its checkpoints create a documentation of this exploration, similar to popular blog posts in this area~\cite{boehm2022optimize}, enabling others to replicate successful workflows and explore similar optimizations in other contexts.

A key aspect of \peak{} is that its fundamental core is general: that is, it does not utilize any backend specific tools, or operate on any specific programming language. Thus, it can be easily ported to different backends. In this work, we implement support for CUDA (NVIDIA GPUs), HIP (AMD GPUs), and HLSL (a portable shading language). However, its extensible interface also allows new tools to be incorporated for specific backends. For example, CUDA has a rich ecosystem of tools, and these could be implemented, e.g., in the correctness validators or performance evaluators. Similarly, branching in performance workflows allow users to write custom transformations that are specific to certain backends, while also sharing more general transformations across backends. 

\paragraph{Immediately Pragmatic and Future Compatible} All of the interfaces provided by \peak{} can be accessed directly by users, e.g., manually or in a script. Furthermore, they also have an MCP interface, so that they can be accessed by natural language during kernel development in an IDE. However, the interfaces can also be driven by an AI agent, creating a completely automated workflow. Thus, \peak{} is designed to be immediately pragmatic, providing productivity benefits for performance engineers today, while also being future-compatible, enabling workflows to become increasingly autonomous as AI technologies evolve.

\subsection{Optimizing Matrix Multiplication and Exploring Research Questions}

Given the low-cost of writing natural transformations, \peak{} can be used in narrow domain specific optimization situations, which is often the case for performance-engineers, who must hyper optimize a small number of kernels. In this style, we focus our evaluation on a single, but critically important, kernel: matrix multiplication (or MatMul). This kernel is a fundamental operation in many high-impact domains where GPUs are deployed at scale, specifically AI. At the same time, it is notoriously difficult to optimize due to the need to balance memory hierarchies, thread utilization and backend-specific features such as tensor cores. As such, kernels are often specialized not only to backend, but to specific devices and even specific input sizes.

Although optimized libraries for MatMul exist (e.g., cuBLAS), they are complex and difficult to adapt or modify. It is useful to have an incremental optimization approach where modifications (e.g., such as those done in kernel fusion) can be incorporated. Furthermore, the knowledge base around MatMul can be useful in more complex kernels that share similar patterns, e.g., convolution and attention. Furthermore, as we illustrate with HLSL, the knowledge base can be used to build libraries for widely deployed GPUs that might otherwise not have a high-performance library. 

We explore both \texttt{fp32} (32-bit floating point) and \texttt{fp16} (16-bit floating point) variants of MatMul and two sizes, 2048 x 2048 and 4096 x 4096 (abbreviated to 2k and 4k). We implement 16 natural transformations, which encapsulate 37 LLM calls. These transformations range from general-purpose GPU optimizations (e.g., loop tiling) to backend- or device-specific strategies (e.g., exploiting tensor cores) and are described in \mysec\ref{sec:transformations}.

\begin{table}[t]
\centering
\small
\caption{Performance summary across devices, backends, and precisions. Speedup is given over the simple input kernel. The two numbers in each column denote values for square matrix of size 2048 \& 4096. The \% Max FLOPS is given as a ratio of the maximum FLOPS we observed in top performing library calls (cuBLAS and hipBLAS for NVIDIA and AMD) or for Qualcomm, over the reported FLOPS for the device. 
}
\label{tab:peak-summary}
\begin{tabular}{@{}lllrrr@{}}
\toprule
\textbf{Device} & \textbf{Backend} & \textbf{Precision} & \textbf{Baseline Speedup} & \textbf{\% of Max FLOPS} & \textbf{Transformations} \\
\hline
\multirow{2}{*}{NVIDIA A6000} & \multirow{2}{*}{CUDA} & \texttt{fp32} & \texttt{9.36 \& 10.36} &  \texttt{95.0 \& 91.6} & 11 \\
                              &                       & \texttt{fp16} & \texttt{41.06 \& 45.34} & \texttt{99.4 \& 89.5} & 9 \\
\hline
\multirow{2}{*}{AMD MI200}    & \multirow{2}{*}{HIP}  & \texttt{fp32} & \texttt{24.39 \&
25.21} & \texttt{91.9
\& 86.2} & 8\\
& & \texttt{fp16} & \texttt{36.14 \& 39.56} & \texttt{48.2 \& 37.2} & 5\\
\hline
Qual. Adreno X1-85 & HLSL & similar & \texttt{4.16
\& \hphantom{1}4.71} & \texttt{107.3 \&
109.8} & 3\\
\hline
\textbf{Average} & & & \texttt{19.52} &  \texttt{78.26} & \texttt{7.3} \\
\hline
\end{tabular}
\end{table}

\paragraph{Evaluation and Research Questions} To evaluate \peak{}, and more broadly to characterize the capabilities of LLMs in GPU kernel optimization, we instantiate a realistic, semi-autonomous optimization scenario. For each GPU and each MatMul variant (\texttt{fp32} and \texttt{fp16}), we provide a sequence from our natural transformations that mirrors how a human performance engineer might plan an optimization strategy. \peak{} then executes this sequence, applying each transformation, validating correctness, and evaluating performance. A summary of the GPUs and their resulting performance is shown in \mytab\ref{tab:peak-summary}. Our final kernels have significant speedup over their baseline inputs and competitive performance with vendor supplied libraries. Surprisingly, although we could not find a standard BLAS library for HLSL, we found that a small number of transformations produced MatMul kernels that achieved slightly over the reported number of FLOPS for this device. This shows the power of \peak{} in transferring knowledge from one backend to another. 

At the time of writing, we note that \peak{} embodies an attractive design choice in this fast-moving area. On one hand, it outperforms many existing on this difficult kernel (matrix multiplication); on the other hand, it does not require a complex RL infrastructure. For example, the KernelBench leaderboard reports MatMul kernels to be around 20\% of library performance. At the time of writing, the AI CUDA Engineer reports their MatMul kernels to be at 42\% of library performance (the public leader has since seemed to be taken down). The CUDA-L2 system reports higher performance on matrix multiplication, but requires a complex RL pipeline and thus, likely many curated examples~\cite{su2025cudal2surpassingcublasperformance}. While PEAK requires lightweight manual prompts, we achieve an average of 78\% of library performance and above 90\% in many cases.
Furthermore, \peak{} can easily be extended with natural transformations once  performance issues are diagnosed, e.g., for AMD \texttt{fp16}, for which there are generally fewer resources optimization strategies and \peak{} currently performs suboptimally.

The human provided sequence of natural transformations is not a limitation of \peak{}. In practice, the transformation space could be searched automatically, or sequences could be proposed by an LLM, potentially guided by a learned or curated knowledge base. Our configuration simply provides a practical and interpretable starting point for systematically exploring the capabilities and limitations of LLM-assisted GPU optimization. Furthermore, this setup enables us to explore a set of research questions at fine-granularity. These questions aim to illuminate both the strengths and current limitations of LLMs in the domain of GPU kernel optimization, helping to identify where existing models are effective and where further research is needed.

\paragraph{Contributions.}

We introduce \peak{}, a performance engineering AI assistant for GPU kernels powered by natural transformations.
To summarize, our contributions are:

\begin{itemize}
    \item The design of a modular and extensible framework built around enabling safe and performant natural language driven transformations for GPU kernels (\mysec\ref{sec:peak-design})
    
    \item An implementation across three GPU backends (CUDA, HIP, and HLSL), demonstrating both cross-backend knowledge transfer and backend-specific specialization. (\mysec\ref{sec:backends}).
    
    \item A case study optimizing MatMul kernels, achieving  competitive performance and optimal GFLOPS for backends without a library (\mysec\ref{sec:workflows}).
    
    \item An investigation of research questions offering insight into the current capabilities and limitations of LLMs for GPU kernel optimization (\mysec\ref{sec:research-questions}).
\end{itemize}


\section{Background: Matrix Multiplication on GPUs\label{sec:background}}
We first review the architecture of modern GPUs using the example of an optimized MatMul kernel. 
Unless otherwise noted, we adopt the terminology used by NVIDIA's CUDA programming model.
\subsection{GPU Programming Model}
Most GPU programming models adopt a split execution model between the \emph{host} (CPU) and the \emph{device} (GPU).
A GPU \emph{kernel} is a function executed by many GPU threads typically following Single Instruction Multiple Data (SIMD) paradigm.
A GPU kernel is written using a GPU programming language, such as, CUDA for NVIDIA GPUs, HIP for AMD GPUs, and HLSL is a portable shading language that can target many different GPUs.
The kernel programming model closely reflects the underlying hardware hierarchy, exposing low-level control needed to implement efficient, high-performance kernels.

GPUs contain multiple Simultaneous Multiprocessors (SMs) to support large-scale parallelism.
Each SM typically contains 64 or 128 CUDA cores, where each core executes one thread.
Data center GPUs contain over 100 SMs while mobile or consumer-class GPUs typically contain between 2 and 20 SMs.
This hierarchical design enables GPUs to scale performance across a wide range of workloads and device classes.

A GPU organizes threads in a three-dimensional grid, where each thread has a three-dimensional index.
The grid of threads is divided into a group of consecutive threads, known as \emph{thread blocks}.
Consecutive threads of a thread block are further divided into fixed-size groups known as a \emph{warp}, where threads execute in a Single Instruction Multiple Threads (SIMT) model.
In a SIMT model, all threads in a warp attempt to execute the same instruction in lockstep; however, when control flow divergence occurs, threads can take different execution paths, reducing the execution efficiency.
The size of a warp depends on the hardware, e.g. 32 for NVIDIA GPUs, 64 for some AMD GPUs, and 128 for Qualcomm GPUs.
The host launches a GPU kernel based on a launch configuration that specifies the number of thread blocks in three dimensions and the number of threads for each thread block.
This three-dimensional layout allows developers to naturally express computations over multidimensional data (e.g., images, matrices, volumes).

A GPU kernel reads its input data and writes its output data to the GPU DRAM, typically referred to as the \emph{global memory}, because it is accessible to all threads of the GPU.
The host is responsible for allocating buffers on the global memory and transferring data from/to these buffers.
The host must wait for the GPU kernel to finish before reading the output from global memory buffers.
A smaller and faster on-chip memory, called \emph{shared memory}, is available to all threads in a thread block.
The shared memory is typically used as a programmer-managed cache to avoid multiple trips to the global memory.
Threads in a thread block can synchronize reads and writes to shared memory using built-in barrier primitives such as \texttt{\_\_syncthreads()}.
Moreover, each thread has access to hundreds of registers, and a thread of a warp can read the registers of other threads of the warp.

Several GPU programming models expose a variety of data types tailored to GPU hardware capabilities.
In addition to standard scalar types like \texttt{float} and \texttt{int}, most GPU backends support low-precision types such as \texttt{fp16} (16-bit floating point) and \texttt{fp8} (8-bit floating point).
Modern GPUs also include \emph{tensor cores} which are specialized hardware units to accelerate low-precision operations and are invoked at the warp level instead of thread level.
GPUs allow batched memory accesses using vectorized types, such as \texttt{float2}, \texttt{float4}, or \texttt{int4}, to improve memory bandwidth.

\subsection{Matrix Multiplication \label{sec:background-matrix-mult}}
Matrix Multiplication (or simply MatMul) is a core computation in many critical and popular workloads.
A standard BLAS library API takes two input matrices \( A \in \mathbb{R}^{\text{m} \times \text{k}} \) and \( B \in \mathbb{R}^{\text{k} \times \text{n}} \) and produces an output matrix \( C \in \mathbb{R}^{\text{m} \times \text{n}} \) such that \( C = A \times B \).
In this work, we assume that all matrices are stored in row-major order and that elements may be stored in either \texttt{fp32} or \texttt{fp16} precision.

A naive MatMul kernel launches one thread per output element \( C[i][j] \), with each thread computing a single dot product between the \( i \)-th row of \( A \) and the \( j \)-th column of \( B \). While functionally correct, this approach fails to exploit memory locality or shared computation, and thus performs suboptimally on modern GPUs.
Several optimizations have been developed to obtain near-maximum throughput provided by GPUs~\cite{cutlass,diesel}.
The standard optimization strategy is to apply \emph{tiling}, where the computation is partitioned into smaller blocks (tiles), across all levels of compute and memory hierarchy of GPUs.

The first level of tiling is \emph{Thread Block Tiling}, where each thread block computes a $t_m \times t_n$ block of \( C \).
Therefore, there are exactly $\frac{m}{t_m}\times \frac{n}{t_n}$ number of thread blocks.
To fully utilize memory reuse and minimize global memory accesses, thread block tiling stores $t_m$ rows of \( A \) and $t_n$ columns of \( B \) in the shared memory.
However, since shared memory size is limited, this tiling also defines a size, $t_k$, such that, each thread block loads $t_k$ elements of both rows and columns into shared memory. 

The next level of tiling, \emph{Warp Tiling}, defines two more tile sizes, $w_m$ and $w_n$, such that there are exactly $\frac{t_m}{w_m}\times \frac{t_n}{w_n}$ warps in a thread block.
A warp processes $w_m$ rows of \( A \) and $w_n$ columns of \( B \) to produce a $w_m \times w_n$ tile of \( C \), which is stored in thread-local registers.
Warp tiling is the last level of tiling for tensor cores for low-precision floating-point operations.

The last level of tiling for CUDA cores, \emph{Thread Tiling}, defines two more tile sizes, $r_m$ and $r_n$, such that $\frac{w_m}{r_m}\times \frac{w_n}{r_n}$ is equal to the warp size of the GPU.
Each thread loads $r_m$ rows of \( A \) and $r_n$ columns of \( B \) into the registers to produce a $r_m\times r_n$ block of C.
Similar to TB-Tiling, this tiling minimizes shared memory accesses by loading input rows and columns to registers from the shared memory.

\subsubsection{Advanced Optimizations}
In addition to choosing where data resides in the memory hierarchy, developers must also consider how memory is accessed between threads.
For example, threads within a warp should ideally access contiguous memory locations to fully utilize the global memory bandwidth.
Similarly, when accessing shared memory, all threads in a warp should access addresses that fall into different banks to avoid shared memory \emph{bank conflicts}.  

Moreover, it is also important to utilize both compute and memory resources of a GPU simultaneously.
To achieve this, efficient MatMul kernels perform \emph{pipelining} in a thread block by overlapping the global to shared memory copy of next $t_k$ with the computation of current $t_k$. 

These optimizations interact in complex ways, and the most performant configuration depends on the specific hardware, precision, and matrix dimensions. Thus, MatMul serves as an ideal case study for evaluating GPU performance engineering and the capabilities of AI-based optimization tools like \peak{}.

\section{Designing \peak{}: An Extensible and Modular Framework for Natural Language-Driven GPU Kernel Optimization \label{sec:peak-design}}

\begin{wrapfigure}{r}{0.55\textwidth} 
  \centering
  \includegraphics[width=1\linewidth]{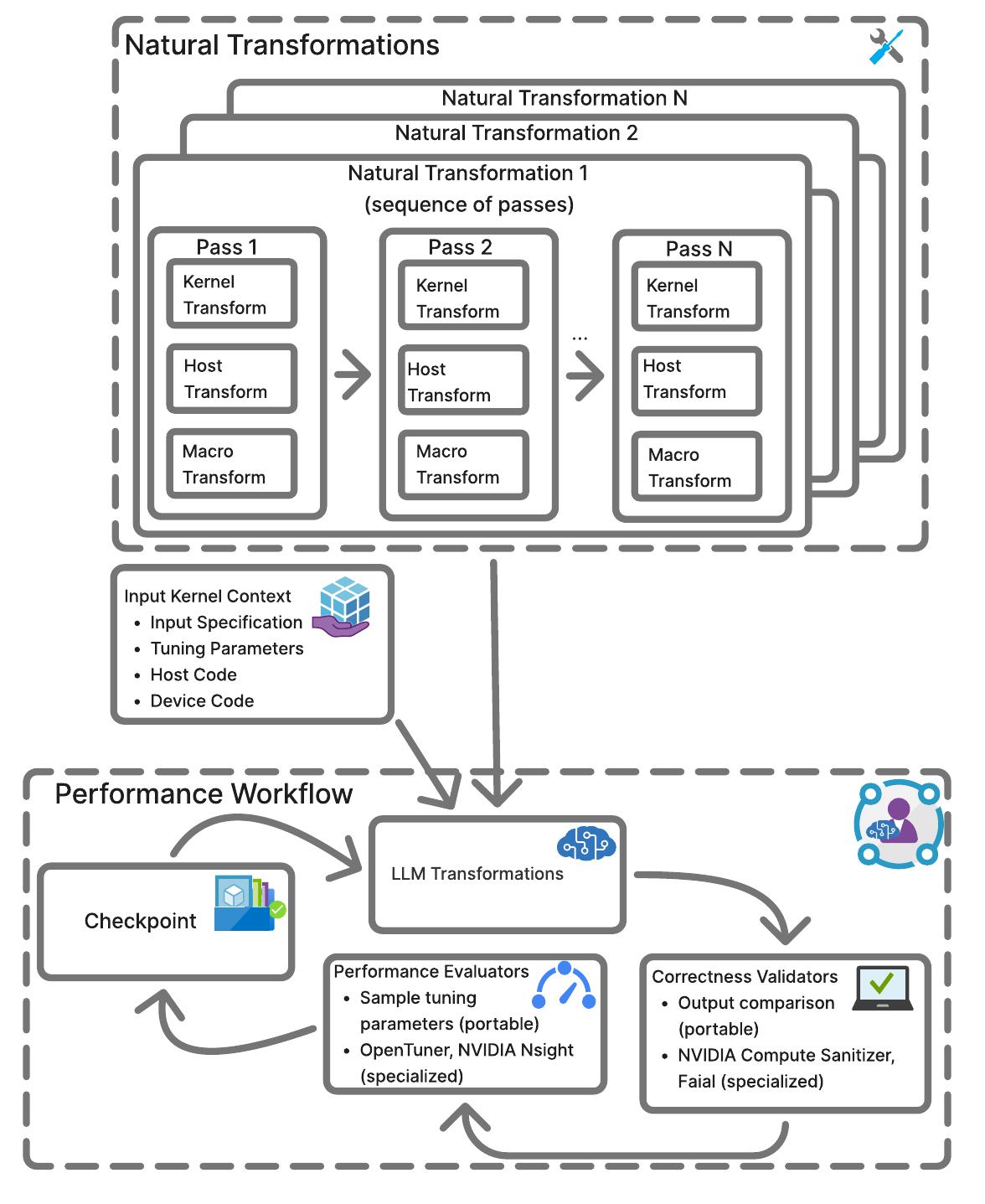}
  \caption{\peak{} design overview}
  \label{fig:peak-design}
\end{wrapfigure}

We now describe the design of \peak{}, illustrated in \myfig\ref{fig:peak-design}, which is centered around lightweight, natural language transformation specifications executed by large language models (LLMs) . While LLMs offer flexibility and generality, the complexity of GPU kernel development necessitates a structured framework to ensure correctness, performance, and reuse. Our high-level design goals are as follows:  
(1) to provide a rigorous scaffolding that enables many simple, natural language transformation specifications, both general and specific;  
(2) to support basic functionality across a wide range of GPU devices and frameworks, which can then be fine-tuned; and  
(3) to enable extensibility, so that researchers and practitioners can incorporate external tools for validation, performance evaluation, and different techniques for kernel transformation, LLM-based or otherwise. We describe the core interfaces of \peak{} and then detail our implementation across three GPU backends: CUDA, HIP, and HLSL.

\subsection{The Kernel Context \label{sec:kernel_context}}

Intuitively, a \emph{kernel context} encapsulates all information required to execute, evaluate, and refine a GPU kernel. This includes:  
(1) the kernel code and any associated device functions;  
(2) a host function that launches the kernel;  
(3) an input specification; and  
(4) a tuning parameter specification.

\paragraph{Input Specification and Performance Tuning Parameters}  
Input arguments are defined using a lightweight specification language. The language supports scalar and array types, including \texttt{f32}, \texttt{f16}, and \texttt{int}; arrays can further be annotated to be \textit{output} arrays, which means their values can be used in the correctness validators. For scalars, all possible values must be enumerated explicitly or using range constructs, akin to list comprehensions in Python. For arrays, the specification must enumerate possible sizes (denoted with the field \texttt{.size}), while the values themselves may be populated using predefined initializations such as zeros, ones, or random values. Constraints may be applied to ensure validity across inputs. For example, if a kernel takes an array \( A \) representing an \( m \times k \) matrix, the specification would enumerate valid values of \( m \) and \( k \), as well as array shapes for \( A \), while using a user-defined \texttt{valid\_args} function to enforce that \texttt{A.size == m * k}.

While the current specification language is not designed to express complex structured inputs, e.g.,  graphs and sparse matrices, it is sufficient to capture a range of impactful GPU kernels, particularly those found in machine learning workloads. In these domains, performance-critical computations are characterized by array shapes and dimensions, rather than specific data values in the arrays. Moreover, performance engineering often involves specializing GPU kernels for specific input shapes. As a result, the space of valid inputs is typically small and tractable, a common assumption in related work~\cite{tvm, halide}. Even production libraries like cuBLAS select different implementations internally depending on input sizes, motivating \peak{}'s support for size specialization.

In addition to input arguments, a kernel context may include \emph{performance tuning parameters}, represented as special string placeholders in the kernel or host code, with possible values stored in metadata. These parameters capture configuration choices that impact performance but not correctness, such as tile sizes and thread block dimensions. Like scalar inputs, tuning parameters are specified using enumerated sets or ranges.
Explicitly incorporating tuning parameters into the abstraction is, to our knowledge, a novel feature of \peak{}, and one that mirrors common practice among human performance engineers. This design encourages transformations to introduce tunable degrees of freedom, which can then be explored automatically and efficiently. It reduces the burden on the LLM to produce highly specific values during code generation, while enabling flexible and generalizable transformations that defer fine-grained tuning to existing mature tools.

Executing a kernel context requires an instantiation of all kernel inputs and tuning parameters, which we call the \emph{execution parameters}. Despite careful specification, not all execution parameters will be valid. For example, a particular tiling configuration may exceed the register or shared memory budget of a GPU. To handle such cases, we allow kernel contexts to return a special value when invalid configurations are detected at runtime; in turn, \peak{} can then simply prune the execution parameters from consideration.

\paragraph{Drivers.}  
A kernel context encapsulates everything needed to generate a lightweight driver program. This driver is responsible for allocating, initializing, and copying inputs to the GPU. It then executes the kernel multiple times to measure runtime performance, reporting an average execution time. Optionally, it can capture and return output arrays, enabling correctness checks.

\paragraph{Limitations.}  
While \peak{} currently operates on individual kernels, we anticipate that future extensions could support a wider context, e.g., considering multiple kernels and memory transfers between them. This would allow \peak{} to be used, e.g., for optimizations like kernel fusion, which combines multiple kernels into one. For now, \peak{} can be applied to kernels that have already undergone fusion via external tools, e.g., TVM~\cite{tvm}. 

\subsection{Natural Language Transformations for GPU Kernels}

The core innovation of \peak{} is its support for \emph{natural language transformation specifications}, or simply \emph{natural transformations}, which describe how to modify one kernel context into another. After a sequence of such transformations, the aim is for the resulting kernel to be significantly more performant than the input while remaining correct. Because transformations are written in natural language and validated by subsequent correctness checks, they can be authored rapidly, need not be fully formalized, and can incorporate both high-level intent and hardware-specific insights. This flexibility enables kernel-specific and device-specific transformations that would be difficult to express in traditional compiler optimization passes.

Despite the general capabilities of modern LLMs, effective transformations still benefit from thoughtful system design. In particular, we decompose transformation tasks both \emph{spatially} and \emph{temporally}, enabling simpler specifications, improved success rates (which we explore in \mysec\ref{sec:rq2}), and more interpretable optimization sequences.

\paragraph{Spatial Decomposition.}  
\peak{} partitions the kernel context into three regions: the \emph{host code}, the \emph{device code}, and \emph{macro definitions}. Each transformation targets only one of these regions. Macros are commonly used in performance-critical GPU code to avoid function call overhead, without relying on compilers to do inlining. Constraining each transformation to a single region reduces the risk of unintended code edits outside the transformation's intended scope, which we observed during early prototyping. For typical GPU kernels, which are short but intricate, this three-way partitioning has proven sufficient. For transforming larger code bases, e.g., as done in Github Copilot~\cite{githubcopilot2025} and Amazon Q Developer~\cite{amazon2025qdeveloper}, more complicated decompositions will be needed.

\paragraph{Temporal Decomposition.}  
Transformations may also be expressed as a sequence of simpler \emph{passes}. Each pass produces a new kernel context, however, it may be incomplete or temporarily incorrect, but it is structured to enable the next pass in the sequence. This decomposition allows complex or multi-part transformations to be specified as smaller, more manageable steps, improving reliability and more precise debugging capabilities.

\paragraph{Manual and Reusable Code.}  
In addition to LLM-guided edits, \peak{} allows transformations to include manual code insertions, such as auxiliary macros or utility functions that do not depend on the surrounding kernel. For example, a transformation may reference a macro that performs efficient global-to-shared memory transfers. Rather than prompting an LLM to generate such helpers, which can be unreliable and inefficient, \peak{} provides a small library of reusable, hand-written utilities that can be imported on demand. One could imagine incorporating more library frameworks into \peak{} transformations, e.g., CUTLASS~\cite{cutlass}, CUTE~\cite{cute}, or Thunder Kittens~\cite{thunderkittens}.

\subsection{Correctness Validators and Performance Evalulators \label{sec:design-correctness-performance}}

While \peak{} is primarily driven by LLM-guided code transformations, its supporting infrastructure provides the scaffolding necessary to ensure correctness and fine-tune performance. This is organized into two extensible interfaces: \textit{correctness validators} and \textit{performance evaluators}. Each contains a general and portable baseline task, with an extensible interface which can incorporate new tools, potentially specialized for a backend, as this field evolves.

Each task takes a kernel context and a parameter that specifies how many execution parameters to sample during evaluation. Exhaustively enumerating all execution parameters is often infeasible, especially when expensive dynamic analysis tools are used, so sampling provides a practical trade-off. We explore this tradeoff more rigorously in \mysec\ref{sec:rq3}.

\paragraph{Correctness Validators}

The correctness validator is responsible for checking the correctness and safety of a given kernel context, which is critical when kernels have been created by LLMs. The portable baseline task simply compares the output arrays of a kernel context against those produced by a reference implementation, although there needs to be some tolerance theshold to account for small floating point discrepancies.

While output comparison is often sufficient, more subtle or nondeterministic bugs, such as data races or memory safety violations, may go undetected without finer-grained analysis. These issues are increasingly important, especially given increasing concerns about GPU security vulnerabilities~\cite{sorensen2024leftoverlocalslisteningllmresponses, gpu-security} and rare, non-deterministic behaviors that can lead to bugs~\cite{asplos15, pldi16}. To this end, \peak{} integrates specialized dynamic analysis tools, such as NVIDIA’s Compute Sanitizer, to detect memory errors and race conditions. These dynamic analysis tools typically require backend-specific integration, and have high runtime costs but can greatly enhance the confidence in the optimization process. Similarly, \peak{} also integrates static data-race detection tools, such as Faial~\cite{Faial}, which, when applicable, provide more coverage (over execution parameters) and execute faster, but might not handle the latest programming features, such as tensor core APIs. Overall, there does not yet seem to exist tool or methods that provide truly formal and rigorous analysis of GPU kernels, and thus, PEAK is designed so that new approaches can be easily incorporated as they are developed. 


\paragraph{Performance Evaluators}

The performance evaluator identifies performant tuning parameters, and optionally collects auxiliary performance metrics to guide further optimization. The portable baseline task enumerates all valid combinations of execution parameters, executes the kernel with each configuration, and records the runtime to identify the best-performing setup.

To scale beyond exhaustive search, the engine integrates with autotuning frameworks, such as OpenTuner~\cite{opentuner}. The profiling engine shares the same task interface model as validation: it accepts a kernel context and a sampling budget, and returns a performance summary over the sampled configurations. In addition to runtime measurements, profiling tasks collect backend-specific performance counters and metrics.
For example, \peak{} integrates NVIDIA's Nsight profiler to extract hardware metrics such as memory bandwidth utilization or shared memory bank conflict rates. These metrics can inform downstream decisions about which transformations to apply next.

\subsection{Performance Workflows}

Having described the core components of \peak{}, we now describe how these elements interact within a complete \emph{performance workflow}. \peak{} supports iterative performance engineering, where a human user or AI agent applies transformations, validates correctness, evaluates performance, and records progress. For human interfaces, we expose a direct API, and also provide an MCP interface so that the tools can be engaged with through natural language. 

\paragraph{Seeding the Workflow.}
The workflow begins with a manually specified \emph{initial kernel context}. In practice, these initial kernel contexts are often easy to write and typically require fewer than 10 lines of code. 
However, while manual authoring is sufficient for many use cases, kernel contexts can also be generated automatically. For instance, related systems, such as KernelBench~\cite{kernelbench} can be used to generate kernels, and then \peak{} can be used to further optimize; this would enable workflows to start from either human-written or AI-generated code.


\paragraph{Validation.}
Validation in \peak{} is grounded by the behavior of the initial kernel context, which is assumed to represent correct functional behavior given that it should be a simple implementation. Upon initialization, this kernel is executed across the input configurations, and its outputs are recorded as a reference. Subsequent kernel contexts, e.g., produced by natural transformations, are validated by comparing their outputs to these references. This validation may be augmented with more sophisticated dynamic analysis tools, but even simple output checks provide a portable and backend-agnostic foundation for correctness checking.

\paragraph{Checkpointing.}
At any point during optimization, the system can create a \emph{checkpoint}, which captures the current kernel context along with any metadata produced by validation or profiling. These checkpoints allow the system to track optimization progress, compare performance across versions, and backtrack after unproductive transformations. Checkpoints also enhance reproducibility and traceability, making it easier to share experiments or resume long-running optimization sessions.


This flexible design enables \peak{} to support a wide range of optimization workflows, from fully automated pipelines to exploratory, human-driven sessions, and positions the framework as both a practical assistant and a research platform for GPU performance engineering.

\section{Optimizing Matrix Multiplication Across GPU Backends \label{sec:backend-and-kernel}}

We now show how \peak{} can be used to optimize MatMul kernels for a variety of devices and GPU programming backends. As discussed in \mysec\ref{sec:background-matrix-mult}, MatMul is a performance-critical operation that is also notoriously difficult to optimize. 

\subsection{Matrix Multiplication Transformations \label{sec:transformations}}

We begin by describing the natural transformations we provide for MatMul based on the strategies described at a high level in \mysec\ref{sec:background-matrix-mult}.
These transformations were developed using a combination of sources, including published documentation (e.g.,~\cite{boehm2022optimize, diesel}), reference implementations (e.g., NVIDIA CUTLASS~\cite{cutlass}), and discussions with experienced GPU performance engineers. In total, we implemented 16 distinct transformations. For brevity, we summarize 9 of them in \mytab\ref{tab:mm_transformations}.

Each transformation is written in a natural language specification.
Many of these transformations are specific to MatMul, for example, explicitly referring to the ``K dimension''.
We do not claim that these transformations are universally applicable across all GPU workloads. However, similar patterns often arise in other compute-intensive kernels, such as convolution and attention, suggesting that many of these transformations could be easily adapted. Furthermore, many are often reusable across GPU backends (CUDA, HIP, and HLSL), as well as data types (e.g., F32 and F16) which we believe represents an important and underexplored axis of portability in performance engineering.

As described in \mysec\ref{sec:kernel_context}, each transformation may consist of a sequence of smaller \emph{passes}, and each pass can invoke the LLM up to three times, once for each code region: host code, device code, and macro definitions. This decomposition enables complex optimizations to be applied incrementally without overwhelming the LLM. Some transformations simply prepare the code for later transformations, for example, Refactor simply moves primitives (like thread IDs) and array accesses into macros, which can then be targetted by later passes. Some passes are relatively complex, like TB-Tiling, requiring 6 passes. Our pilot experiments found that LLMs struggled with performing this transformation all at once, and breaking it down into simpler steps was more reliable. Indeed, prior work on compilers that perform tiling shows how transformations like this can be broken down into simpler steps~\cite{halide}. 
We found LLMs were able to perform other types of tiling (i.e., across warps and threads) more reliably, and thus, needed less decomposition.

\begin{table*}[t]
\footnotesize
\centering
\caption{Nine out of 16 of our \peak{}'s natural transformations for MatMul}
\label{tab:mm_transformations}
\begin{tabular}{@{}l p{6.2cm} c c@{}}
\toprule
\textbf{Name} & \textbf{Description} & \textbf{Passes / LLM Calls} & \textbf{Tuning Params} \\
\midrule

Refactor & Refactors primitives and complex accesses to macros & 2 / 3 & None \\

TB-Tiling & Performs thread block tiling & 6 / 9 & \texttt{TILE\_K\_SIZE} \\

Warp-Tiling & Performs warp tiling & 1 / 1 & \texttt{WARP\_X\_DIM} \\

Thread-Tiling & Performs thread tiling & 3 / 3 & \texttt{TRD\_X\_DIM} \texttt{TRD\_Y\_DIM} \\

Tensor-Core & Utilizes tensor cores & 1 / 1 & None \\

Split-K & Splits the K dimension across thread blocks & 1 / 3 & \texttt{K\_SPLITS} \\

Transpose & Transposes memory when tiling & 1 / 1 & None \\

Offset & Offsets shared memory tiles to avoid bank conflicts & 1 / 2 & \texttt{OFFSET\_AMOUNT} \\

Pipelining & Pipelines loading and computing thread block tiles & 1 / 2 & \texttt{NUM\_STAGES} \\
\bottomrule
\end{tabular}
\end{table*}

\subsection{GPU Backend Implementations}
\label{sec:backends}

A key strength of \peak{} is that its core design is entirely backend-agnostic: it contains no framework-specific grammars, parsers, and very little hardcoded tooling. This makes it straightforward to instantiate \peak{} for different GPU programming frameworks, which has often posed considerable difficulty for other frameworks, and they provide limited portability. To demonstrate this, we implement \peak{} for three widely used backends: CUDA, HIP, and HLSL.

The most backend-specific component is the driver logic for each kernel context. While the device kernel specifications is similar across backends, the host-side code used to manage memory and launch kernels varies. For example, HIP and CUDA require different API calls for memory management and synchronization, while HLSL, being rooted in graphics APIs, requires significantly more boilerplate, including explicit creation of device handles, command queues, and memory views. Similarly, there is a small amount of hard-coded code in some transformations, mostly related to efficiently loading memory. We were able port this code from CUDA to HLSL straightforwardly.

Beyond this, only a small number of transformations required backend-specific adaptation. Many differences (such as how thread identifiers are provided) were handled by factoring out backend-specific syntax and adding translation tables directly into the transformation prompts. For example, thread indexing syntax differs across CUDA (\texttt{threadIdx.x}), HIP (\texttt{hipThreadIdx\_x}), and HLSL (\texttt{SV\_DispatchThreadID}). A small number of transformations required new specifications; for example, in the case of tensor core integration, the HIP backend required a dedicated transformation due to API differences in naming and supported tile shapes. However, this new transformation was quickly adapted from the CUDA version taking less than 30 minutes that were largely spent in consulting the HIP documentation. 

As noted in \mysec\ref{sec:design-correctness-performance}, \peak{} contains portable validation and performance tasks, such as output comparison and tuning parameter exploration. Given the accessible tooling for CUDA, we were able to provide some specialized tooling. We incorporated OpenTuner for CUDA and HIP, but were unable to use this tool for HLSL on Windows; instead since only few transformations were applied, we could simply enumerate and search over all parameters.

\subsection{Performance Workflows \label{sec:workflows}}

\begin{wraptable}{l}{0.46\textwidth}
\footnotesize
\centering
\caption{GPU devices used in our evaluation and their documented peak throughput (in TFLOPS), separated into \texttt{fp32} / \texttt{fp16} when available.}
\label{tab:devices}
\begin{tabular}{@{}l l l c@{}}
\toprule
\textbf{Device} & \textbf{Driver} & \textbf{Compiler} & \textbf{TFLOPS} \\
\midrule
N. A6000  & 575.57.08  & \texttt{nvcc 12.2} & 38.7 / 155 \\
A. MI200        & ROCm 6.3.2      & \texttt{hipcc 6.3} & 47.9 / 383 \\
Q. X1-85 & 31.0.112.0 & DXC 6.6  & 4.6 \\
\bottomrule
\end{tabular}
\end{wraptable}

We now describe the \peak{} performance workflow used to optimize variants of MatMul across different devices and GPU programming frameworks. The input kernel context is a simple 6-line GPU implementation in which each thread computes a single element of the output matrix \( C = A \times B \). The kernel accepts four arguments: three arrays (\texttt{A}, \texttt{B}, and \texttt{C}), and an integer \texttt{n} representing the matrix dimension. All matrices are square of size \( n \times n \), and array \texttt{C} is designated as the output; all of them are initialized with random values. For this case study, we evaluate two matrix sizes: \( n = 2048 \) and \( n = 4096 \) referred to as 2K and 4K, respectively. We consider two data types: \texttt{fp32} and \texttt{fp16}. 
Our initial kernel context includes tuning parameters that control how the global   thread grid is partitioned into thread blocks. We restrict the thread block dimensions to powers of two, constrained by backend-specific limits (e.g., 1024 for CUDA and 256 for HLSL). 

\begin{table*}[t]
\footnotesize
\centering
\caption{Transformation sequences per device and precision.  The final performance is given in \mytab\ref{tab:peak-summary}. 
}
\label{tab:opt_sequences}
\begin{tabular}{@{}l l p{8.5cm} c@{}}
\toprule
\textbf{Device} & \textbf{Precision} & \textbf{Transformation Sequence} & \textbf{\# Transforms}  \\
\hline
A6000 & \texttt{fp32} & 
Refactor $\rightarrow$ TB-Tiling $\rightarrow$ Thread-Tiling $\rightarrow$ Thread-Cache $\rightarrow$ Transpose $\rightarrow$ Thread-Chunk $\rightarrow$ Split-K $\rightarrow$ Pipelining $\rightarrow$ Register-Staging $\rightarrow$ Offset $\rightarrow$ Block-Swizzle &
11 \\
\hline
A6000 & \texttt{fp16} & 
Refactor $\rightarrow$ TB-Tiling $\rightarrow$ Warp-Tiling $\rightarrow$ Tensor-Core $\rightarrow$ Tensor-Tiling $\rightarrow$ Pipelining $\rightarrow$  Register-Staging $\rightarrow$ Offset $\rightarrow$ Block-Swizzle &
\texttt{9} \\
\hline
MI200 & \texttt{fp32} & 
Refactor $\rightarrow$ TB-Tiling $\rightarrow$ Warp-Tiling $\rightarrow$ Tensor-Core $\rightarrow$ Tensor-Tiling &
\texttt{5} \\
\hline
MI200 & \texttt{fp16} & 
Refactor $\rightarrow$ TB-Tiling $\rightarrow$ Warp-Tiling $\rightarrow$ Tensor-Core $\rightarrow$ Tensor-Tiling $\rightarrow$ Offset $\rightarrow$  Block-Swizzle $\rightarrow$ Pipelining &
\texttt{8}  \\
\hline
X1\mbox{-}85  & both & 
Refactor $\rightarrow$ TB-Tiling $\rightarrow$ Thread-Tiling &
\texttt{3}  \\
\hline
\end{tabular}
\end{table*}

We study the optimization trajectories of \peak{} across three devices, as listed in Table~\ref{tab:devices}. For each device and precision variant, we combined documentation and discussions with GPU performance engineers to construct a transformation sequence designed to yield high-performance kernels. These sequences were assembled from \peak{}'s natural transformations, many of which are given in \mytab\ref{tab:mm_transformations}, and the complete optimization paths for each configuration are summarized in \mytab\ref{tab:opt_sequences}.

While transformation sequences were constructed manually for this study, they could also be discovered via alternative methods, such as automated search or LLM-driven agent-based exploration. However, we believe that this semi-autonomous setting, with a human guiding and interpreting the optimization trajectory, provides a compelling and realistic use case. To support future work in more autonomous optimization, we evaluate how performance evolves throughout these transformation workflows in \mysec\ref{sec:rq1}. 

While many optimization sequences begin similarly, e.g., refactoring and performing thread block tiling, they diverge depending on architectural features. For instance, on NVIDIA GPUs, the \texttt{fp16} path transitions to using tensor cores while \texttt{fp32} uses thread tiling but both variants then converge again at pipelining.
On AMD MI200, both \texttt{fp32} and \texttt{fp16} kernels utilize tensor cores; however, the effective tuning parameters differ between the two, as found by our performance evaluators. For example, the best-performing warp tile size for \texttt{fp32} was \texttt{(4,2)}, while \texttt{fp16} was not found to utilize tiles. Unlike on NVIDIA devices, we did not apply pipelining transformations for AMD on \texttt{fp32}, as the MI200 offers less shared memory, which is a critical resource for this optimization.
%
%
%
For the HLSL backend on the Qualcomm Adreno GPU, only a few transformations were needed to reach the device’s reported peak GFLOPS. Given this, and our focus on other devices and components, we did not explore deeper optimization sequences for this backend.

\begin{wraptable}{l}{0.55\textwidth}
\centering
\footnotesize
\caption{End-to-end time for the A6000 performance workflow. Times are reported as in seconds, as \emph{raw (percent)}. In both cases, the total time is less than 6 hours, which can be run overnight.}
\label{tab:perf_workflow_time}
\begin{tabular}{@{}l r r r r@{}}
\toprule
\textbf{Type} &  \multicolumn{1}{c}{\textbf{Validation}} &  \multicolumn{1}{c}{\textbf{Transforms}} & \multicolumn{1}{c}{\textbf{Performance}} & \textbf{Total} \\
\midrule
\texttt{fp32} & 825 (3.91\%) & 1285 (6.09\%) & 18995 (89.99\%) & 21105 \\
\texttt{fp16} & 1124 (1.55\%)    & 257 (6.77\%) & 15226 (91.68\%) & 16607 \\
\addlinespace[0.25em]
\bottomrule
\end{tabular}
\end{wraptable}

For the sequences reported in \mytab\ref{tab:opt_sequences}, the LLM was able to correctly apply all the transformations in the sequence without human intervention using the o4-mini LLM; however we show in \mysec\ref{sec:eval:llm-performance} that other models are able to successfully (and reliably) apply our transformations as well. 

The end-to-end time to perform all transformations varies greatly on how \peak{} is invoked. For example, how many execution parameters are sampled at each transformation to validate correctness and gather profiling information, which ultimately may just be collected at the end of a performance workflow. Similarly, it is difficult to provide reliable measurements from LLM calls, as we have found their latency varies significantly seemingly randomly (likely based on cloud availability). Thus, with these combined considerations, we provide a summary of the end-to-end execution time of fp16 and fp32 on the A6000 in \mytab\ref{tab:perf_workflow_time} (Our other performance workflows were executed in a more distributed and exploratory manner for the results in \mysec\ref{sec:research-questions}). For the NVIDIA workflows, we sampled 16 execution parameters after each transformation to validate, using both output comparison and NVIDIA's compute sanitizer. After every transformation, we used the performance evaluator to enumerate all performance parameters, and prune all except for the top 128 to utilize in the next transformation. We utilized 128 CPU cores to compile kernels in parallel and 4 GPUs on the machine to execute performance and correctness tasks in parallel. We note that the tuning parameter exploration is the vast majority of the runtime. If future approaches, e.g., symbolic performance estimations, could accurately prune this search space effectively it could improve \peak{} significantly.

\section{Research Questions: Exploring LLMs for GPU Kernel Optimization \label{sec:research-questions}}

Unlike monolithic or opaque approaches that treat optimization as a black box, \peak{}'s structured workflow provides visibility into each transformation step, validation outcome, and performance change. This transparency enables systematic exploration of LLM capabilities and limitations in this domain. We investigated three key research questions that illuminate both the strengths and current limitations of LLM-assisted GPU optimization.

\subsection{RQ1: How Does Performance Evolve During Iterative Optimization Sequences?}\label{sec:rq1}

\begin{figure}
    \centering
    \begin{minipage}[b]{0.48\textwidth}
        \centering
        \includegraphics[width=\textwidth]{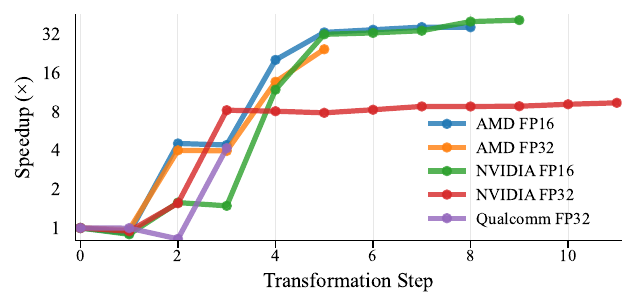}
        \vspace{-2em}
        \caption{Speedup of each transformation over the \emph{naive input kernel context}.}
        \label{fig:sampleimage441}
    \end{minipage} \hfill
    \begin{minipage}[b]{0.48\textwidth}
        \centering
        \includegraphics[width=\textwidth]{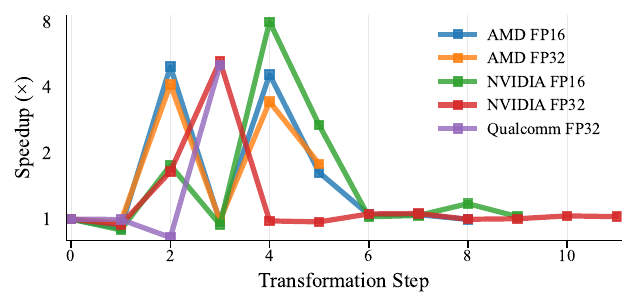}
        \vspace{-2em}
        \caption{Speedup of each transformation over the \emph{previous transformation}.}
        \label{fig:sampleimage442}
    \end{minipage}
\end{figure}

Understanding the performance trajectory during optimization is crucial for developing effective automated strategies and managing expectations about optimization potential. 

\paragraph{Methodology}
We analyze the performance evolution across our five device-precision combinations (AMD FP16/FP32, NVIDIA FP16/FP32, Qualcomm FP32) at size 2K through two complementary views: cumulative speedup relative to the baseline kernel (\myfig\ref{fig:sampleimage441}) and step-wise improvement between consecutive transformations (\myfig\ref{fig:sampleimage442}). The executed steps for each correspond to the sequences given in \mytab\ref{tab:opt_sequences} and has been tuned to find efficient performance parameters. Note that the different instances have different lengths because they have a different number of transformations. 

\paragraph{Key Observations}

\myfig\ref{fig:sampleimage441} reveals that optimization follows distinct phases rather than smooth, incremental progress. Most configurations exhibit a characteristic pattern: minimal improvement in early steps (transformations 0-2), followed by rapid acceleration during critical steps (transformations 3-4), and finally plateau behavior in later stages (transformations 5-10), where small improvements occur over many transformations. For example on NVIDIA \texttt{fp32}, while the red line appears flat starting at step 3, in reality, this long journey provides a 10\% performance improvement. This follows the common sentiment that most effort is spent on the last percentages.

The non-linear performance evolution also suggests that early transformations unlock the potential for subsequent optimizations. That is, the very early transformations, such as refactoring, establishes code structure, but provides limited immediate performance benefit. The acceleration phase (with the highest jumps) appears to coincide with memory hierarchy optimizations and specialized accelerator usage (e.g., shared memory usage, register usage, tensor cores), while the final transformations are more about fine-tuning access patterns.

The magnitude of final speedups varies across device-precision combinations. NVIDIA FP16 and AMD FP16 achieve the highest speedups (35-40$\times$), while NVIDIA FP32 shows more modest gains (8-10$\times$). Qualcomm FP32 demonstrates the shortest optimization trajectory, reaching its performance ceiling after only three transformations. The precision-dependent speedup patterns reflect the underlying hardware capabilities and optimization opportunities. FP16 configurations achieve higher speedups partly due to tensor core utilization and increased memory bandwidth efficiency, while FP32 optimizations are more limited.

The early saturation observed in Qualcomm FP32 suggests that this platform reaches peak performance with fewer transformations, possibly due to architectural differences or more limited optimization opportunities compared to data center GPUs. This is likely due to much less parallelism and absence of dedicated GPU memory in the SoC.

\myfig\ref{fig:sampleimage442} provides insight into which individual transformations drive the most significant improvements. The highest speedups occur at transformation steps 2-4 across most configurations, with individual transformations providing up to 8$\times$ improvement over the previous one.

\subsection{RQ2: How Do Different LLMs Perform at GPU Kernel Transformations? \label{sec:rq2}}
\label{sec:eval:llm-performance}
\begin{figure}[t]
    \centering
    \includegraphics[width=\textwidth]{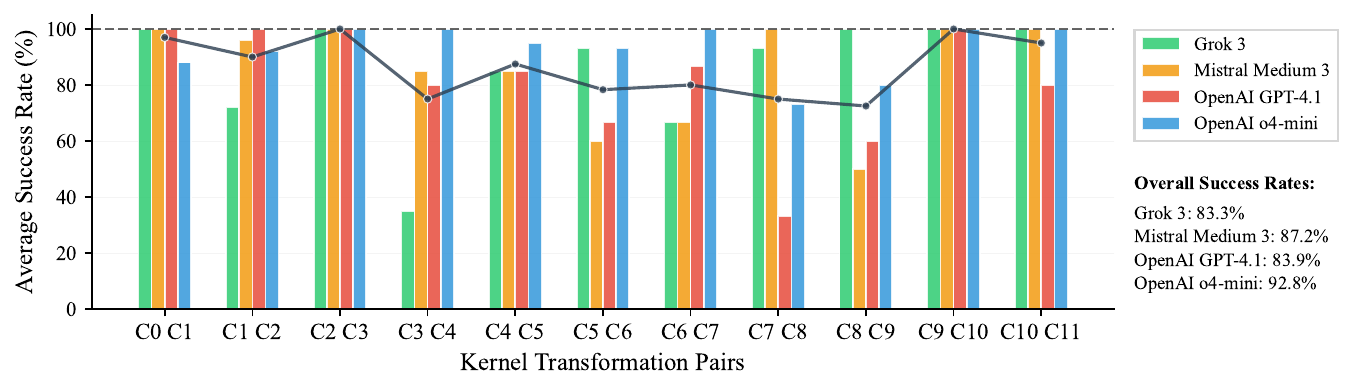}
    \caption{Success rate of each model across transformation pairs. The transformations are done for our three devices - NVIDIA A6000, AMD MI200, and Qualcomm. The figure also presents the overall success rate for each model and also the average success rate for each transformation across all models.}
    \label{fig:model-success-rates-rq2}
\end{figure}

The choice of LLM for executing natural transformations impacts the success rate of individual transformations and, thus the overall optimization workflow reliability. Understanding which models excel at GPU kernel transformations, and which transformation types pose the greatest challenges, can guide practitioners in choosing between different LLM providers.

\paragraph{Methodology}
We evaluate four contemporary LLMs across our complete set of transformation pairs (C0 → C1 through C10 → C11 given in \mytab\ref{tab:opt_sequences}); we run each transformation five times across each model and measure success rates. Success is defined as generating syntactically and functionally correct code, as checked by our validation engine (by sampling 16 execution parameters and performing output comparisons). Note that transformations may not be same across device-precision combinations, as noted in \mytab\ref{tab:opt_sequences}, instead the point is to evaluate how well LLMs perform across a performance workflow. 

\paragraph{Key Observations}

Our results are summarized in \myfig\ref{fig:model-success-rates-rq2}. At a high-level, we see that \emph{all} models are able to successfully apply all transformation steps, even if they do not have 100\% success rate. This means, any model can successfully run \peak{}, if given enough attempts at each transformation. We see this as a success of our transformation decomposition into a series of simpler passes.

Despite this, there are still performance differences between models, with overall success rates ranging from 83.3\% (Grok 3) to 92.8\% (OpenAI o4-mini). The two reasoning models (OpenAI o4-mini and mistral medium 3) are the top two performers, which highlights that reasoning models perform slightly better in this domain. We note that we utilized o4-mini in many of our pilot experiments, and thus, we may have inadvertently biased our prompts to work better with that model.  

The transformation-specific analysis reveals that certain optimizations pose challenges while others are consistently well-handled across models. The success rates also seem to go in phases, with early and late transformations being performed well, but difficulty in the middle transformations, i.e., C3 → C4 up to C8 → C9. We believe this highlights the difficulties of the tasks: early transformations (thread block tiling, thread tiling, tensor cores) modify the code extensively, but they are typically well-understood (i.e., there are many examples) and follow regular patterns. Middle transformations tend to modify the code extensively, have fewer examples, and be less regular. For example, we found pipelining, transpose, and register staging (where global memory values are staged in registers before being stored to shared memory) to be particularly difficult for LLMs. Then later passes, while conceptually more complex, are highly targeted, and modify the code relatively little compared to earlier passes.

These error rates could likely be improved in several ways: (1) by fine-tuning the prompts to each model, or (2) by breaking down the transformations into even simpler passes. Given that \peak{} also supports human interaction, it is possible for the LLM to implement \emph{most} of a transformation, and for a human to finish editing the kernel context, and then create a checkpoint, and move on to later transformations through the LLM, which tend to be more reliably executed.

\begin{figure}[t]
    \centering
    \includegraphics[width=\textwidth]{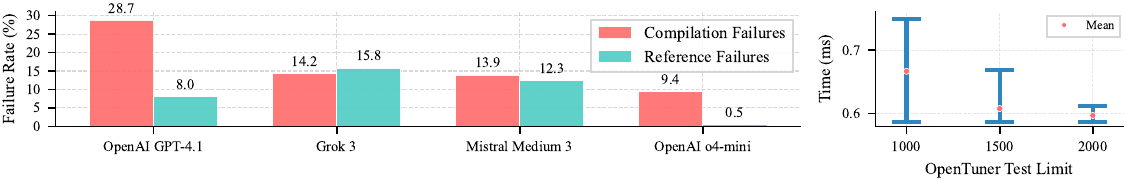}
    \caption{Left: average failure rate for each model across all transformation pairs. Compile errors are counted as Compilation failures whereas kernels that compile but do not produce the correct output are counted as Reference failures. The transformations were done for our three devices - NVIDIA A6000, AMD MI200, and Qualcomm. Right: Time range achieved by OpenTuner for the final Matrix Multiplication kernel (FP32) on AMD MI200. Higher test limit means that OpenTuner was  run for more iterations.}
    \label{fig:rq3}
\end{figure}

\subsection{RQ3: To What Degree Does Correctness and Performance Need to be Evaluated?}\label{sec:rq3}

Understanding the depth of evaluation (both for correctness and performance) required for different LLMs and hardware configurations is important for designing efficient optimization workflows. While comprehensive evaluation provides confidence, excessive evaluation can become computationally expensive (even in our example workflow shown in \mytab\ref{tab:perf_workflow_time}, performance evaluation consumes 90\% of the time). The challenge lies in determining the optimal balance between thoroughness and computational efficiency for different system configurations.

We analyze two dimensions of evaluation requirements: (1) the types and frequency of errors produced by different LLMs, categorizing failures into compilation errors and reference failures, i.e., whether the output array values match the reference, and (2) the performance evaluation depth required, examining how different OpenTuner test limits affect performance discovery across hardware backends.

\paragraph{Key Observations on Error Patterns.} The left graph of \myfig\ref{fig:rq3} reveals variation in error types across LLMs. OpenAI GPT-4.1 exhibits the highest overall failure rate (28.7\%), with compilation failures dominating over reference failures. In contrast, OpenAI o4-mini shows the lowest failure rate (9.9\%), with compilation failures representing the majority of its errors (9.4\% vs 0.5\% reference failures). Grok 3 and Mistral Medium 3 show more balanced error distributions, with reference failures comprising a substantial portion of their total failures.

The dominance of compilation failures across most models suggests that syntactic correctness remains a primary challenge for LLMs in GPU kernel optimization. However, we see this as an opportunity, because typically compiler errors produce actionable error messages, which LLMs can use to recover. However, reference failures are far more worrisome. They take longer to catch, as the code must be compiled and executed. It is also unclear if the errors are deterministic (e.g., caused by a data race) or if they occur for all execution parameters, or if they require extensive searching. Finally, it is difficult to get actionable feedback that can be fed back into the LLM (however, this would be a promising direction for future work). Thus, we see strong potential in using models that have low reference failures; in our case, we found that o4-mini has significantly fewer reference errors (0.5\%).

\paragraph{Key Observations on Performance Evaluation.} The right graph of \myfig\ref{fig:rq3} shows the varation of runtimes found by OpenTuner when run on the AMD GPU for the final kernel of \texttt{fp32} on size 2K. We run OpenTuner 5 times for the three different iteration counts shown on the x axis. The graph shows the distribution of the times found across these 5 executions, showing the maximum, minimum, and mean. We see that extending test limits from 1000 to 2000 iterations (from roughly 10 to 20 minutes) shows continued performance improvements, with mean execution time decreasing and variance reducing substantially. This suggests that AMD hardware benefits from more extensive parameter exploration, likely due to complex interactions between tile sizes, memory hierarchies, and compute unit utilization.
Conversely, analysis of NVIDIA fp32 configurations (not shown) indicated faster convergence, with results often converging in just several hundred iterations. This may be due to a more regular and predictable performance profile on NVIDIA GPUs, as OpenTuner implements advanced searching mechanisms that aim to exploit patterns. Regardless, this disparity highlights the hardware-specific nature of performance optimization and the need for adaptive evaluation strategies.

\subsection{Putting it all Together}

\peak{}'s design enabled exploring these three research questions in a precise and fine-grained manner, which provides concrete takeaways for any AI assistant for performance optimization. In particular, we highlight the following insights:

\begin{itemize}
    \item Given the insights into RQ1, the performance differences between transformations may not always be smooth or even monotonically increasing. Thus, when searching for optimized kernels in a black box manner, similar to~\cite{AICUDAEngineer}, there may need to be more sophisticated evaluation measures, or the ability to explore deeply without expecting immediate benefits.
    \item Given the insights into RQ2, models tend to perform well at early and late stages in optimization workflows but struggle in the middle. Thus, explorations should spend extra time in these stages; potentially providing finer-grained prompts, or enabling extra retries. 
    \item Given the insights into RQ3, models should be evaluated on the types of errors they produce and validation tasks should be adjusted accordingly. Regardless, there should be a feedback mechanism to repair compiler errors. For performance, different devices should be calibrated in order to determine how much exploration is needed. This is especially important as this evaluation time can be very costly, as shown in \mytab\ref{tab:perf_workflow_time}.
\end{itemize}

\section{Related Work}\label{sec:relatedwork}

\paragraph{Optimizing GPU Kernels} 
Several works have aimed to improve the developer experience in writing optimized GPU kernels.
There are several higher-level domain specific languages (DSLs) for optimizing image processing and machine learning applications on GPUs, like Halide~\cite{halide}, Exo~\cite{exo}, PolyMage~\cite{polymage}, and CoCoNet~\cite{coconet}.
These DSLs express a computation on a tensor as a stage, then enable a set of transformations on stages like fusion, reorder, overlapping, and generate optimized code for various GPU backends. REPTILE~\cite{reptile} and TACO~\cite{taco} similarly provide scheduling abstractions for tiling strategies and sparse tensor operations respectively.
Triton~\cite{tillet2021triton} and TileLang~\cite{tilelang:2025} are higher-level kernel languages that abstracts intricacies of the hardware and its low-level kernel language.
Both of these languages provide a tile based abstraction, where the programmer writes a vectorized code for processing a tile and the tile is mapped to one or more thread blocks at runtime.
Similarly, FlexAttention~\cite{flexattention:2025} is a higher level abstraction for generating various kinds of attention kernels.
Lower-level intermediate representations languages, like TVM~\cite{tvm}, TensorIR~\cite{tensorir}, and Graphene~\cite{graphene}, apply several transformations to generate efficient GPU kernels.
Unlike DSLs, ThunderKittens~\cite{thunderkittens} is an extensive library that can be used to write CUDA kernels using warp-level primitives like tensor cores and efficiently manage memory transfers and pipelining.
In contrast, \peak{} enables quickly writing efficient GPU kernels using existing kernel languages.
Moreover, we believe \peak{} is extensible enough to utilize the transformations of the above compilers as natural language text to generate other optimized GPU kernels.
We leave this study for future work.

\paragraph{AI Assisted Optimizations} New advances in AI have enabled LLM-based approaches for code optimizations. For example, LLM Compiler~\cite{llm_compiler:2025} trains on large set of LLVM IR and assembly to emulate optimizing compiler passes, achieving nearly the same performance. LLM-Vectorizer~\cite{llmvec:2025} applies an LLM to loop vectorization on LLVM IR and integrates the formal verifier Alive2~\cite{alive2:2021} to validate the transformed code. While these works show that LLMs can implement optimizations, they are either one-shot systems or implement opaque search strategies. They lack support for interpretable iterative refinement, rollback, or human-in-the-loop guidance, which \peak{} provides. Other approaches were surveyed in~\cite{survey_llm_opt:2025}, showing trends across prompt-based, fine-tuned, and reinforcement-driven methods, while noting challenges such as limited real-world evaluation, lack of correctness guarantees, and difficulty integrating into production toolchains. This emphasizes the difficulty of utilizing LLMs in this domain and highlights the importance of \peak{}, as it can explore LLM performance in a fine-grained manner, as we do in \mysec\ref{sec:research-questions}.

\paragraph{AI Optimizations for GPU Kernels} As mentioned throughout, there are several works that utilize LLMs to optimize GPU kernels. KernelBench~\cite{kernelbench} provides an extensive corpus of AI tasks for GPU programs, from individual kernels, to entire DNNs. Their current leaderboard shows that LLMs can produce functional kernels in many cases, but complex kernels like MatMul, only achieve a fraction of the performance as vendor libraries. The GPU Kernel Scientist~\cite{gpu_kernel_scientist:2025} and the CUDA Engineer~\cite{AICUDAEngineer} performs iterative search over kernel transformations, where an LLM repeatedly mutates CUDA or HIP kernels using runtime performance feedback. While this approach is also iterative, there is little insight into the search, and as we show in \mysec\ref{sec:rq1} the performance workflow may contain extended paths where there is little performance gain, and even performance loss occasionally. As a result, the current results for the AI CUDA Engineer shows only around 45\% performance of vendor libraries for MatMul, whereas \peak{}, requiring some manual work in writing prompts, can achieve over 90\% in many cases and will only improve as it's curated knowledge base grows. Recent concurrent work has explored similar directions. Hong et al.~\cite{hong2024llmaided} investigate LLM-aided compilation for tensor accelerators. Zhou et al.~\cite{zhou2025qimeng} propose QiMeng-GEMM, which is very similar to PEAK, in which specialized prompts are used to optimize GEMM kernels, similar to the natural transformations in PEAK. While similar in spirit, PEAK provides additional experiments showing the fine-grained capabilities of LLMs in this domain. Furthermore, PEAK explores more diverse input and backend domains, considering FP16 (requiring tensor cores transformations) and shows that prompts can be generalized across different vendors, including AMD and Qualcomm.

We note that this is a fast moving area, and as mentioned in the introduction, there are now several start ups also making significant progress in this area, e.g., see~\cite{Stanley2025_Mako, su2025cudal2surpassingcublasperformance, standardkernel_2025}. We believe that PEAK offers an attractive design point in being interpretable, i.e., allowing inspection and development at every stage of optimization, and accessible, i.e., not requiring a complex RL infrastructure, while also producing competitive performance for important kernels.

\section{Conclusion}\label{sec:conclusion}

\peak{} utilizes advances in AI technology to transform GPU kernel performance engineering from an ad hoc, expert-driven task into a transparent and reproducible process that combines human expertise with AI-driven automation. Rather than attempting monolithic and opaque end-to-end generation, it applies iterative natural-language transformations, with the ability to check correctness and fine tune performance in a portable, yet extensible, manner. Our evaluation on matrix multiplication shows that \peak{} outperforms other work in the area, and even achieves competitive performance with vendor-tuned libraries on NVIDIA and AMD GPUs. Furthermore, the curated knowledge base developed for one GPU can transfer to entirely different backends, creating high performance kernels where libraries might not even exist, as we illustrate for HLSL. 


\bibliographystyle{ACM-Reference-Format}
\bibliography{sample-base}

\end{document}